\documentclass[runningheads]{llncs}
\usepackage{graphicx}
\begin{document}
\title{Constructing Dreams using Generative AI}
%
%
%
\author{Safinah Ali\and
Daniella DiPaola \and
Randi Williams \and
Prerna Ravi \and
Cynthia Breazeal}
\authorrunning{S. Ali et al.}
%
\institute{Massachusetts Institute of Technology, Cambridge MA 02143, USA }
\maketitle              
\begin{abstract}
Generative AI tools introduce new and accessible forms of media creation for youth. They also raise ethical concerns about the generation of fake media, data protection, privacy and ownership of AI-generated art. Since generative AI is already being used in products used by youth, it is critical that they understand how these tools work and how they can be used or misused. In this work, we facilitated students’ generative AI learning through expression of their imagined future identities. We designed a learning workshop - Dreaming with AI - where students learned about the inner workings of generative AI tools, used text-to-image generation algorithms to create their imaged future dreams, reflected on the potential benefits and harms of generative AI tools and voiced their opinions about policies for the use of these tools in classrooms. In this paper, we present the learning activities and experiences of 34 high school students who engaged in our workshops. Students reached creative learning objectives by using prompt engineering to create their future dreams, gained technical knowledge by learning the abilities, limitations, text-visual mappings and applications of generative AI, and identified most potential societal benefits and harms of generative AI.

\end{abstract}
\section{Introduction}
Generative Artificial Intelligence (AI) algorithms are types of machine learning models that are used to create novel data samples that are similar to examples it was trained on. These algorithms have been used to create a variety of media such as text, images, videos, 3D models, and music, and have found applications in art, imaging, engineering, protein folding and modeling~\cite{aggarwal2021generative,rombach2022high}. Rapid expansion in their capabilities have led to generative AI algorithms making their way into consumer tools such as OpenAI’s ChatGPT, which reported to have surpassed 100 million users within five months of first release~\cite{milmo2023chatgpt}. There has also been a rapid reduction in the barriers to access these tools. Tools like Stability AI’s Dream Studio\footnote{https://beta.dreamstudio.ai/} or OpenAI’s DallE-2\footnote{https://openai.com/dall-e-2} made text-guided image generation available to users with little technical knowledge or complex computational resources. Media generated using AI and features supported by generative AI have also made their way to social media platforms such as TikTok, Instagram, Reddit and Twitter, which are frequently used by high school students~\cite{abi2020smartphones}.

Generative AI algorithms reduce barriers to creative expression and introduce a novel medium, opening up new creative possibilities. Contrastingly, generative AI tools also harbor several ethical concerns, such as ownership and copyright, data security, plagiarism, generation of fake information and the spread of misinformation~\cite{rezwana2022identifying}. They can have an adverse effect on creative careers, and have been shown to reproduce harmful biases against groups already historically marginalized by technology. Although generative AI tools and media are commonplace in technologies used by youth, and have direct implications for their education, communities and future careers, previous work has demonstrated how K-12 students have little understanding of how these tools work~\cite{druga2017hey,ali2021exploring}. As the use of generative AI tools proliferates, it becomes increasingly imperative to educate youth about them. There have been very few efforts around designing curricula and learning materials for K-12 generative AI literacy~\cite{ali2021exploring}. 

Making in creative media is a powerful means of self-expression for youth, which in turn supports their technical learning~\cite{kafai2011youth}. Given the creative nature of generative algorithms and their ability to support children’s creative expression, it is apt to use them to support both self-expression and technical learning. Expression of self-identity and narratives has been central to creative and STEM learning pedagogy~\cite{hawkins2002children,taylor2005self}. In this work, we designed an informal learning workshop that employs (1) constructionism as a means of learning the technological capabilities of generative AI algorithms, (2) expression of self-identity and narratives for future as a means of enabling creative expression and sharing, and (3) critical debate as a means for reflecting on the societal and ethical implications of generative algorithms. 

We designed the workshop “Constructing Dreams”, where high schoolers imagined a future dream that they visualize for themselves., and used a text-to-image generative tool called Dream Studio by Stability AI to visualize their dreams. Through this informal activity, we aimed to achieve the following learning objectives: 

\subsection{Creative learning objectives:}
C1. Students create input prompts appropriate for generating images using a generative AI tool. \\
C2. Students iteratively tweak their creations to match their creative goals. 

\subsection{Technical learning objectives:}
T1. Students are aware of the abilities of text-to-image generation algorithms. \\
T2. Students identified limitations in the abilities of text-to-image generation algorithms.\\
T3. Students identify visual features in a generated image mapped to parts of textual prompts.\\
T4. Students map patterns in generated images to patterns in underlying datasets used to train the algorithm. \\
T5. Students identify areas of application of generative algorithms.

\subsection{Ethical learning objectives:}
E1. Students identify potential harms of generative AI: algorithmic bias, copyright infringement, creation of fake media leading to the spread of misinformation, plagiarism, job replacement and data privacy.\\
E2. Students voice their opinions about the school policy around the use of generative AI in classrooms. 

We conducted the activity with two student groups (N=13 and N=21). In this paper, we share the design of the activity, students’ creations and their responses that allude to the aforementioned learning objectives.

\section{Background} 
\subsection{Generative AI}
For the purpose of this work, we refer to data-driven generative machine learning algorithms as Generative AI. Generative AI algorithms learn from patterns of existing data (also known as training data), and generate novel data in the form of images, videos, audio, text, and 3D models. While generative models have long existed, recent years have seen a rapid expansion in the field of data-driven generative AI with the introduction of GANs, VAEs and Diffusion models. Algorithms like Dall-E, that received both text-image pairings as inputs, coupled image generation algorithms with large language models and enabled text-guided image generation~\cite{ramesh2021zero}. Tools developed by OpenAI such as ChatGPT or Dall-E2, and subsequently by Stability AI~\footnote{https://stability.ai/} or Midjourney~\footnote{https://www.midjourney.com/} made generative AI accessible to everyone with internet access. There has been a significant discourse about generative AI algorithms’ potential impact on students’ learning in classrooms and future careers~\cite{baidoo2023education}. On one hand, generative AI tools have the potential to give young creators powerful new tools to express their ideas in different mediums. On the other hand, some worry that over-reliance on generative tools may hinder creativity in the classroom. Moreover, generative AI tools harm visual artists by using their work without their consent or affecting their future prospects, are accompanied by data protection and privacy concerns, can potentially lead to misinformation and contain biases that amplify stereotypes. Still, K-12 students do not have a complete understanding of how these everyday algorithms work~\cite{ali2021gans}. We believe that it is imperative, now more than ever, for high school students to learn about what generative algorithms are, how they actually work, what they can be used for and what their ethical and societal implications can be. 

\subsection{K-12 AI Literacy and constructionism}
Traditional AI learning requires familiarity with mathematical concepts and programming skills that K-12 students often lack. The AI4K12 national initiative established five big ideas in K-12 AI literacy that encompass what youth must know about AI~\cite{touretzky2019envisioning}. Curricula, learning toolkits and activities have since emerged that aim to teach children about the 5 big ideas, several of them focusing on creative applications of AI~\cite{zhou2020designing}). Previous work used Block-based coding environments that have enabled students to train supervised learning algorithms and use them in projects~\cite{williams2022ai+,jordan2021poseblocks,kahn2018ai}. Researchers developed K-12 AI education curricula that emphasize constructionist learning, designing with ethics in mind, and developing a creative mindset~\cite{ali2019constructionism}. Previous work also used learning trajectories for teaching middle school students about generative models where they explored the technical constitution, practical applications and ethical implications of generative algorithms such as GANs through unplugged activities, creative exercises and ethical reflections~\cite{ali2021exploring}. Students learned to generate text and images using generative algorithms and could articulate benefits and harms of generative AI for our society. Since then, generative AI tools have become more affordable, user-friendly, and accessible over time. Furthermore, their outputs have become higher quality, leading to more widespread application of generative AI tools in various fields, including creative arts, business, and scientific research. 

In this work, we learn from the constructionist learning approach of previous AI curricula that use AI project building via digital and physical toolkits as means to teach AI concepts, as well as their focus on developing students’ ethical and creative abilities. Constructionist learning approach also emphasizes the importance of engaging in conversation with one’s own and others’ artifacts~\cite{papert1991situating}. We use a sharing and discussion approach, where students reflect on generative AI through their and their peers’ generated artifacts. In addition to existing approaches, we also leverage expression of self-identity as a means for allowing students to create with and learn about generative AI. 
 
\subsection{Identity construction in technical education}
Previous research has identified technical education as a powerful avenue for identity construction. \cite{umaschi2001identity} developed identity construction environments (ICEs) as technological tools designed with the goal of supporting young people in the exploration of self-identity by engaging learners in the design of a graphical virtual city and its social organization. Studies have explored the role of identity in learning coding through game-making~\cite{kafai2015constructionist}. \cite{pinkard2017digital} used narrative driven curriculum to spark non-dominant girls’ interests in STEM activities and identification with the discipline. \cite{bers1998interactive} also developed Storytelling Agent Generation Environment (SAGE), an authoring environment where children created programmable storytellers that allowed them to explore and present their identities. They demonstrated how, through the process of building their own storytellers, and exploring and communicating identity issues, children developed structured thinking required for programming and debugging. Within AI learning, \cite{zhang2022integrating} explored representation of identity and stereotypes to teach middle schoolers about racial bias in classification machine learning algorithms. 

Identity representation exercises become especially relevant avenues of generative AI learning, because students go in with a visual expectation of how they want to represent themselves, and aim to create a language description. Liu et al. (2022) explore how prompt engineering plays a major role in creators’ abilities to use Text-to-Image Generative Models. Creators are met with a generation that could be close to or far from their original expectation, and make correlations with how AI represents words in their natural language description. They may draw mappings of which visual features correlated with which words - e.g. “in Disney style” got represented as colorful wide strokes and friendly cartoons, which related concepts got represented - e.g. “sinking” was related to an unmentioned water environment, which words were considered more prominent e.g. “a writer, painter and a cobbler” only visually represented a painter, and what stereotypes and biases exist in the algorithm, e.g. “a doctor” got represented as a White man. 

In this work, we build upon previous work using identity construction for technical learning by using a future identity construction and sharing exercise using generative AI with the goal of teaching high school students about its technical and ethical implications. Representation of identities of others, particularly of non-white people in western popular culture have been riddled with harmful stereotypes~\cite{hall1997spectacle,zhang2010asian}. Stereotypical representation of groups underrepresented in computing, such as women, has shown to be a barrier to inclusion for members of the group~\cite{cheryan2013stereotypical}.  We chose an exercise involving depicting future selves, so students gain power over constructing their own representations, intentionally overcoming stereotypes and potentially visualizing anti-stereotypes that aid their sense of belongingness in technical fields. Sharing their dreams through textual and visual representations was also a means of creative storytelling and sharing aspirations with their peers. 

\section{Methods} 
\subsection{Participants}
We recruited high school students from two independent high-school STEM programs that provide students with informal learning opportunities: [name redacted] in the US and [name redacted] in India. Students voluntarily signed up to participate in our workshops, it was not part of their academic requirements. Workshop 1 (W-1) was conducted with 17 students in person at [university name redacted] and workshop 2 (W-2) was conducted remotely with 21 students located in [city name redacted], India. As part of their participation in the workshop, students could also take part in our research, but they were not obligated to. In W-1, 13 out of 17 students consented for their data to be recorded for research purposes, and in W-2, all students provided informed consent for their responses to be recorded. Responses from students who did not consent were neither recorded, nor presented in this paper. Participant demographic is presented in table-1. All students were enrolled in high school, and we did not collect information about their age.

\begin{table}
\caption{Students recruited for the workshop}\label{tab1}
\begin{tabular}{|l|l|l|l|}
\hline
Workshop &  N (with consent) & Location & Gender\\
\hline
Workshop-1 (W-1) &  13 & [redracted], US & F = 7, M = 4, Unknown = 2 \\
Workshop-2 (W-2) &  21 & remote, India & F = 11, M = 10 \\
\hline
\end{tabular}
\end{table}

\subsection{Workshop Design}
The workshop duration was 1 hour and 20 minutes. We deliberately made our workshop a single session so that it could be used either as a quick, informal introduction to AI or as part of longer AI or art curricula. The workshop was divided in the following segments: 

\subsubsection{Introduction to Generative AI}
Students were first asked whether they were familiar with generative AI tools, followed by whether they were familiar with ChatGPT. A researcher with expertise in AI introduced students to the concept of generative AI using the following three methods: 

\textbf{Definition}. The researcher described generative AI algorithms as types of algorithms that learn from patterns of existing data (also known as training data), and generate novel data in the form of images, videos, audio, text, and 3D models

\textbf{Activating (and contrasting with) prior knowledge}. The researcher contrasted generative AI algorithms with AI algorithms that predict, such as classification algorithms. While predictive algorithms are trained on many images of cats and dogs and accomplish tasks like classifying between a cat and a dog, generative algorithms are trained on many images of a cat to create a new image of a cat that doesn’t exist, but looks like a real cat.  

\textbf{Showing examples}. The researcher displayed some examples of generated media such as images, text, video, audio and 3D models. 

\textbf{Using Generative AI Tools}. The researcher proceeded to explain to students how generative AI algorithms can be used using images, videos or audio of the following examples: (1) Generating lesson plans with ChatGPT. (2) Generating visual art in a particular artist’s style with Diffusion models. (3) Collaborating with a music generation algorithm to co-create music. (4) Creating videos from movie scripts. (5) Creating novel protein structures that bind to insulin receptors. (6) Creating concepts for an advertisement campaign. (7) Creating Deepfakes - or generating manipulated videos or audio of people. The examples were chosen such that they cover a variety of generative media and show examples of both positive and negative use cases. 

\subsubsection{Practicing reflecting on implications}
With the aim of helping students to be ready to reflect on the ethical implications of AI systems and be aware of their limitations as they engage with the generative tools, the researcher used guided reflection for an example generated artwork. The researcher first displayed examples of AI-generated artwork and asked students how such a technology can be used positively, followed by how it can be used negatively. Students were then asked who they think should get the monetary benefit from selling the artwork. Further, the researcher displayed Deepfakes and asked students how such a technology can be used negatively, followed by how it can be used positively.  Finally, the researchers displayed some examples of generations that demonstrate algorithmic bias in generative AI tools. 

\subsubsection{Constructing Dreams}
In this part of the activity, students were first asked to imagine a dream in words - “If you could be anything in your future, go anywhere, do anything, what would you be doing? It could be a place or an activity, or a job that you see yourself in that does not currently exist. Write your dream in as much detail as you can. Prompt writing was scaffolded using a combination of the following questions and example responses: 
\begin{itemize}
\item What do you look like in this dream?
\item Where are you located in this dream? 
\item What are you doing that represents this dream? Think of job roles that you don’t see yourself represented in today, or create one that does not exist today.
\item What are some visual features that you would like to represent this dream?
This combined story serves as a prompt for the text-to-image generation algorithm. 
\end{itemize}
 
Students use text-to-image the generation tool Dream Studio that uses Stable Diffusion to generate images in different visual styles from a text prompt. Dream Studio was used because of its impressive creation abilities, variety of styles, ability to disable uploading of content and their checks around inappropriate prompts and content. Creators can also add a negative prompt of components they wish to omit from their generation. In order to prevent students from adding personal information, we provided them with a shared account linked to a research ID and disabled the ability to upload students’ images on the website. Further, students were made aware of Dream Studio’s privacy policy around ownership of content - “you own the Content that you generate using the Services to the extent permitted by applicable law and Stability and our affiliates may use the Content to develop and improve the Services, including by storing your Content and associated metadata”. Students were asked to edit their prompts until they were satisfied with the generated image and posted them to a shared Google Slides document. 

\subsubsection{Sharing Dreams}
In addition to generating their dreams, an essential part of the learning activity was to share their dreams with the class, providing an avenue to express their imagined future identities and investigate AI’s perceptions of their expressed identities. The researcher displayed the dream images and corresponding prompts that students shared, and called upon students to share their creations and discussing them using the following prompts: 
\begin{itemize}
\item What were your first impressions of the tool? Were you surprised by the creations? 
\item How did you think AI represented different parts of your prompts? 
\item Were you satisfied with the generations? If not, what did you change?
\item How were the generations different from your expectations? 
\item What were some useful tricks you learned? 
\end{itemize}

\subsubsection{Reflection}
After the creative activity, students reflected on generative AI as a class, guided by the following prompts presented by the researcher: 
\begin{itemize}
\item What are some ways you think generative models can be useful to you?
\item What do you think are some positive uses of generative AI models? 
\item What are some negative ways you think generative AI models can be used? 
\item Do you think the jobs that people do will change due to generative AI? 
\item How can generative AI be used in the classrooms? 
\item Do you think using generative AI to do your assignments be permitted in classrooms? Do you think there should be any conditions on using them? 
\end{itemize}

We did not conduct a formal assessment from every student due to time constraints and the informal structure of the workshop, but two researchers recorded participants’ informal classroom discussions, which are summarized in the results below.

\section{Results} 
\subsection{Introduction to Generative AI: Knowing it when you see it}
At the beginning of the workshop, we asked students whether they were familiar with generative AI, and only 6 out of 34 students responded yes. However, 29 students reported being familiar with OpenAI’s ChatGPT. Students were engaged in the introductory explanation and examples of generative AI and found examples of realistic generative media surprising. We found no apparent differences in engagement between the two groups.

\subsection{Initial reflections: Seeing the potential benefits and harms}
Among the benefits of generative AI, students expressed making art, making movies, and sharing your ideas with others. One student in W-2 mentioned, “when I have difficulty explaining what I am thinking to someone else, I can just use this.” Among the harms, one student in W-1 mentioned that “it is not really art and essentially cheating from real artists.” Another student called generating art in real artists’ style “plagiarism.” A student in W-1 mentioned how “using ChatGPT to write an essay would be cheating.” Another student in W-2 mentioned that “it could take up artists’ jobs if we can just generate art for free.” 

Students expressed variation in who they think should benefit monetarily from the sale of AI-generated art. In W-1, one student mentioned how it should be the “coder who made the algorithm”. Another student expressed how they think such a painting should not be sold at all, “[I think it] shouldn’t be used or sold at all, because it’s taking from other peoples art without getting their permission, it feels like stealing– who should be giving permission”. Several students brought up stealing of artworks from artists without their permission. One student mentioned how the money should be split between all the people involved in the process, but was unsure how the split should be decided. In W-2, a majority of students expressed that the company or person that “designed the AI” should get the money from the artwork, and added that “they should pay the artists for the data.” 

While discussing potential harms of Deepfakes, one student in W-1 said that Deepfakes can be used to “Impersonate people in negative ways, make someone say something bad - this might be met with panic and aggression from the community.” Another student said that it can lead to the “spread of misinformation and fake news.” Another student highlighted that they make the credibility of real information poor, “Today there are a lot of people that believe Obama is a Deep Fake - he doesn’t exist– makes you question real information altogether.” In W-2, students discussed how it can be used to “run smear campaigns or create false evidence in courts.” Among potential benefits, students in W-1 identified how they can “speak to ancestors who have passed away” or “speak to their younger versions.”.  In W-2, a student said how they can create their own entertainment by making movie characters they like act in different ways. 

\subsection{Constructing dreams: Expressing imaginations through generative AI}
\begin{figure}
\includegraphics[width=0.9\textwidth]{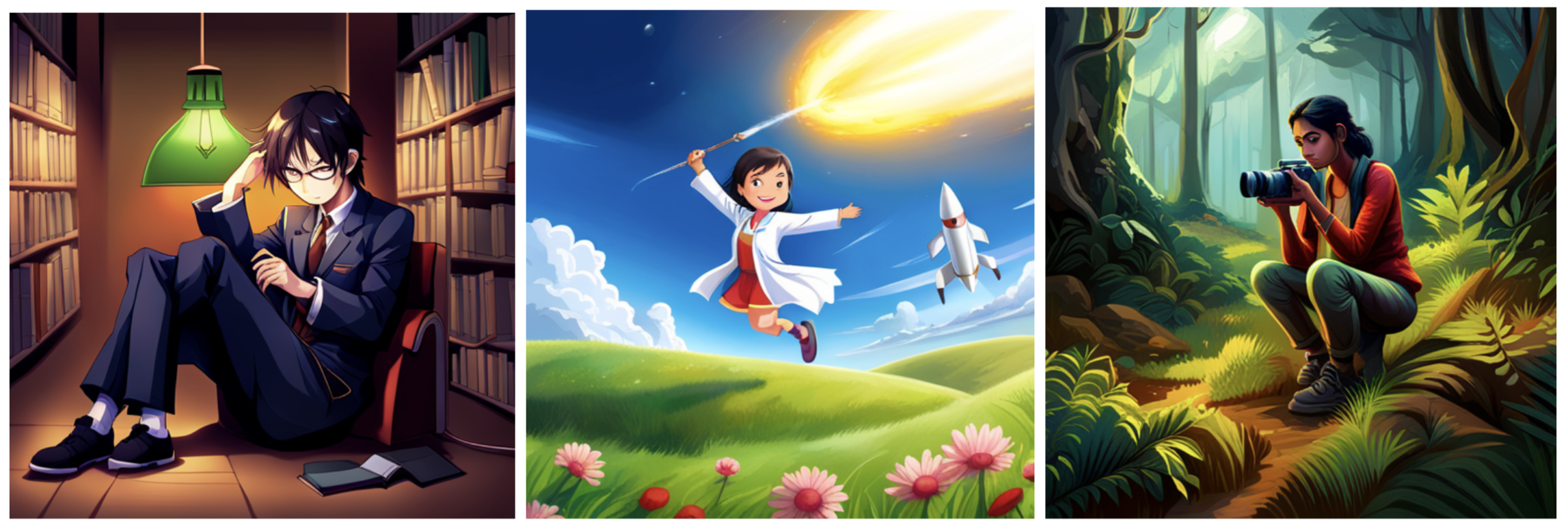}
\caption{Three students’ generated dreams using the prompts: a. “A Vietnamese male wearing a dark casual academia outfit doing research on psychology in a library with a lot of books.”, b. “A young and small asian girl with black hair and fair but not white skin wearing an extra large lab coat with a rocket in one hand that is about to launch while jumping and wearing a large smile on her face, A large open space with blue skies and green grass and little wind. One tree in the background and a dog that’s running along, Launching a small homemade rocket that is about to fly in one hand as a part of an experiment.” c. “An indian girl who is a photographer in the forest nature, with a camera, shooting a documentary, wearing cargo pants, high detail in illustration style”. } \label{fig1}
\end{figure}

All students were able to generate dream images using their text prompts. Students generated an average of 12.6 images (min: 4, max: 31), indicating that all students tried multiple attempts to reach their desired image. All students finished with a prompt longer than their initial prompt. Students tweaked their positive prompt, negative prompt, style options and number of images. All students started with the scaffolding questions provided, but eventually deviated from those to add and remove details that better suited their goal. Figure~\ref{fig1} depicts generated dreams created by three participants along with their prompts. Through the generating dreams activity, we could read the creative learning goals, where students created input prompts appropriate for generating images using a generative AI tool (C1) and iteratively tweak their creations to match their creative goals (C2).

\subsection{Sharing Dreams: Exploring the art of prompt engineering}
Only 26 out of 34 (W-1: 8, W-2: 18) students shared final generated images they were satisfied with. One student mentioned that the generated image was “Highly accurate, I liked the first one that I saw.” Whereas another student said, “I had to play around with the details with the things that I emphasized and the styles, every picture has a different vibe to it.” Multiple students reported dissonance with what was first generated and their desired images. For instance, one student generated using the prompt, “A hispanic boy with curly hair in a minion suit” and the tool generated a minion face with curly brown hair, but he expected a boy with a minion space suit. He kept tweaking the prompt and adding more details to get to an image he desired, and ended with the prompt, “A brown hispanic boy long curly hair glasses wearing a minion suit. located in space on my own planet with land I found with other fellow minions. my own mechanic and building my own ship with my robot friend i built. visual features high detail and 4k”. 

All students who shared out reported having to add more details to bring their image to a desired point. Students added details to their appearance and the setting to better reflect aspects of the dream, such as attire to depict a chosen profession. Nine students also chose to write a negative prompt to indicate features they wanted to omit. Students also shared useful tips with their peers. One student mentioned how using his age in the prompt helped him get closer to the desired image. Another student used the name of a famous person who they believed they resembled. One student added visual information to their prompt, “visual features high detail and 4k”. One student said that they found it useful to add what they want in the background and remove information about their dreams that were “not related to the looks”. All students in W-2 identified as South Asian, and were trying to represent a similar ethnicity, which is not the default representation in Dream Studio. Students came up with, and shared their representation strategies, such as “using Hindi words” or say “South Indian and specify skin tone” or “say Pakistani instead of Indian so it wouldn’t confuse with Native Americans.” 

Students also shared examples of algorithmic bias and stereotypes in generative AI. One student mentioned how, while trying to depict a South Asian face, when they said “Asian” the image primarily depicted an East Asian person, and saying “Indian” would either depict a Native American person or did not resemble them. The student shared that adding words of Indian attire, such as, “lehenga” (a colorful Indian long skirt) helped them get closer to the desired depiction. Another student commented on how “pretty girl” would depict a young White face and said, “it kept making me blonde”. The instructor prompted students to probe why this bias exists, and one student correctly pointed out how the data used to train these algorithms have such stereotypes.

Through examples in the introduction session and an interactive activity, students were made aware of abilities of text-to-image generation algorithms (T1). Pertaining to the learning objective T2, students identified the following limitations in the abilities of text-to-image generation algorithms: stereotyping and algorithmic bias, mismatch in creators’ and AI’s representation of concepts and a feeling of lack of agency. Students also developed prompt engineering skills to reach their desired image. Through sharing prompt tips and identifying assumptions that the algorithm made around textual prompts, students could identify visual features mapped to textual prompts (T3). Students themselves did not make an explicit connection to patterns in underlying training datasets (T4), but upon instructor’s probing about why they think “pretty girl” is represented as a White blonde young girl, students could identify that, “the examples it is using has more pretty girls labeled as White.”  Students related the knowledge from the introduction session of how generative AI uses examples of text-image pairs to predict visual patterns associated with text. 

\subsection{Reflection: Imagining AI in the classroom and beyond}
Students began reflecting by thinking about how generative AI models can be useful to them (Objective T5). One student from W-1 said how it can be used in “projects in school to visualize future selves”. Another student said they can “study anatomy to practice art by visualizing organs.” A student from W-2 expressed how they can use it for their arts class to “visualize paint styles”. One student recognized that a positive use of generative AI can be to “help you explain things, you can collaborate with others best”. Among negative uses, students expressed how it can “Fake images and make things very convincing”. Students also discussed how jobs will change as a result of generative AI tools. One student in W-1 believed that creative jobs will not alter - “I don’t think so because AI art still has a far way to go, some of it doesn't look good and it doesn't look human level yet.” Another student pointed out that medical diagnoses can change. Another student in W-1 said that they “don’t think teachers’ jobs could change at all, teachers need to know different types of children.” One student in W-1 remarked how “AI cannot be emotional.” and another in W-2 said that “AI will never understand people as people do.” Students overall did not believe that creative careers will change. However, one student in W-1 said that they believe that writing jobs such as journalism or creative writing, storytelling could be affected by it. 

Within the classroom, students in W-1 discussed how it can be used to “make presentations more visually appealing” or for “adding visuals to a book for visual learners.”  Students in W-2 discussed how history could be taught using imagining visuals from the events. When students were asked about whether generative AI should be permitted in classrooms, one student from W-1 responded that its use should be allowed, but when generative AI was used to complete a task, it should be noted as such. Another student from W-1 said that it could be used in classrooms, but not in homework as there will be no way to tell. Students in W-2 overwhelmingly responded that using generative AI should not be permitted in classrooms. 

Through the sharing and reflection activity, we could partially reach ethical learning objective E1, where students identified the harms: algorithmic bias, creation of fake media leading to the spread of misinformation, plagiarism and job replacement. Copyright infringement was not identified as such but was referred to as “stealing” media. Students did not identify data privacy as a potential harm. Beyond the stated objectives, students identified the loss of credibility of real information as a potential harm. Some students voiced their opinions about the school policy around the use of generative AI in classrooms (E2).

\subsection{Discussion}
Our learning activity “Dreaming with AI” helped students express their future identities using generative AI, share their experiences with the tool with their peers and reflect on the ethical and societal implications of generative AI. Students successfully used a text-to-image generation tool to express their dreams and share them with their peers. Students performed “prompt engineering” and discovered and shared tricks to aid their generations. Students could reach all creative learning objectives. Among technical learning objectives, students did not map patterns in generated images to patterns in underlying datasets used to train the algorithm themselves (T4), but could make that connection upon further probing by the instructor. Students could not identify data privacy among potential harms of generative algorithms (E1). In future work, the interactive activity could involve investigating components of generated images that map to parts of text prompt. Students could also reflect on the sources of datasets used for training generative algorithms and who owns the data to emphasize data privacy.

Using an identity expression activity followed by reflection was a useful approach to teaching students about generative AI for several reasons. Firstly, students were able to visually express and share their dreams with their peers, allowing them to experience a generative AI tool’s powerful creative abilities. In students’ visualizations we saw them creating representations of themselves that countered stereotypical depictions of their gender, age and racial identities. Second, students directly experienced and confronted the limitations of the generative tools. Students stumbled upon examples of bias displayed by algorithms in the form of generative tools reflecting a version of the world that was different from students’ own mental images. Through probing and reflection, students engaged with the biases inherent in the datasets and media representations used to create these tools. Finally, students could connect these experiences back to the potential use cases of generative AI and voice more nuanced opinions on the appropriateness of using these technologies in the classroom. For students to make the most of this experience it was important that they were primed with knowledge about how generative AI algorithms function and how they may make errors.

We conducted this workshop with two groups: in the US and in India. The workshop itself did not require any adaptation in content since most examples chosen were globally known. A difference in results observed between the two groups was that the students in the US were a lot more positive about using generative AI in classrooms as compared to students in India who primarily viewed it as plagiarism. Previous research has shown how students in higher education in Hong Kong revealed a generally positive attitude about generative AI, while also recognizing some ethical concerns~\cite{chan2023students}. More work needs to be done to understand what leads to these cultural differences. Another difference we noted was that since W-2 was more a more racially homogenous group, while also sharing unique struggles to generate an Indian identity, they were able to share prompt tricks with their peers to get their desired representation. 

We encourage AI or arts educators to adapt this workshop to teach high school students about generative AI tools, their applications, and their ethical implications.

\subsection{Limitations and Future Work}
This work presents an early exploratory workshop geared towards introducing high school students to generative AI tools. Results present students’ responses to activities and reflection prompts, but since no formal assessment was conducted and we did not have every students’ responses, inferences about all students’ learning cannot be made. A longer workshop with formal assessments of students’ understanding of AI concepts is required to make inferences about the efficacy of learning materials. Future work can aim to capture more detailed insights about students’ knowledge about generative AI prior to the workshop, embed assessments around the creative process and decision making in generating dreams, conduct a more formal assessment of AI knowledge and attitudes with all students and understand cultural differences between creators from different geographic locations.

\subsection{Ethical considerations}
Generative AI tools recreate harmful stereotypes in datasets, and must be brought to young learners responsibly with educator supervision. AI tools are often incorrect with believable confidence. Children could possibly overestimate or develop misconceptions about the abilities of AI, or not have the ability to discern real data from generative media. It is important to discuss the limitations of the tools with students before interacting with them. Technologies such as generative AI are socio-technical systems that have positive and negative impacts on society and the workforce. It is important to have conversations with students about the capabilities and potential harms that these tools may cause. 

\bibliographystyle{splncs04}
\bibliography{paper}

%
%

\end{document}